# Selección de zonas de interés en el Valle de Aburrá y San Nicolás con el método de identificación de clústeres basados en densidad y vecino cercano mejorado aplicado en redes sociales.

*Selecting Areas of Interest in Valle de Aburrá and San Nicolás Town Using the Identification Method of Density-Based Clustering and Improved nearest Neighbor Applied in Social Networks.*

Esteban Zapata Rojas[*]

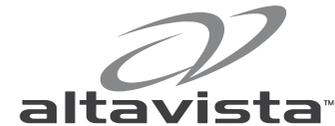

**Resumen**

Más que nunca, las redes sociales han cobrado un importante puesto en la interacción y el comportamiento de los humanos en la última década. Esta valiosa posición hace que sea imperativo analizar diferentes aspectos de la vida cotidiana y la ciencia en general. En el presente artículo se ilustra cómo fue el proceso de captura y almacenamiento de la información, la aplicación de los métodos de Clustering basados en densidad y vecino cercano mejorado en los datos y un repaso por los resultados, mostrando los elementos utilizados en la identificación de las zonas de interés a través de clústeres, circunferencias y radios de cobertura obtenidos para realizar un análisis de segmentación demográfico de la información obtenida a través de las redes sociales Twitter y Flickr, permitiendo conclusiones más profundas sobre un tema o temática predefinida. Por último, se plantea la necesidad de elaborar una aplicación que automatice todos los procesos acá definidos, permitiendo que usuarios finales interesados en estos temas tengan acceso a ella y puedan obtener resultados importantes para su organización o interés.

**Palabras clave:** Clúster, redes sociales, identificación de clústeres, machine learning.

**Abstract**

More than ever, social networks have become an important place in the interaction and behaviour of humans in the last decade. This valuable position makes it imperative to analyze different aspects of everyday life and science in general. This paper illustrates the process of capturing and storing information, the application of density-based clustering and improved nearest neighbor, and a review of the results. The study also shows the elements used in the identification of areas of interest through clusters, circumferences and coverage radii obtained for a demographic segmentation analysis of the information procured from Twitter, Flickr and the like. This results in more profound conclusions about a predefined topic or theme. Finally, the need arises to develop an application that makes all the defined process automatic, allowing final users interested in those topics to have access to it and get important results for their organizations or interest.

---

* Desarrollador en Altavista Editores. Miembro del colectivo de comunicación y redes sociales CaféSM. ezapata@altavistaed.com.

# Introducción

Las redes sociales se han convertido en el centro de atención de millones de personas alrededor del mundo en la última década y cada año crecen de manera exponencial, teniendo casos interesantes como el de Twitter. (April & June, 2014). En Colombia, el uso de estas tecnologías se ha masificado al punto de ser una de las 15 naciones con mayor crecimiento en el uso de redes sociales en el año 2012 (Semiocast, 2012) y un crecimiento del 278 % en el número de dispositivos móviles activos entre enero del 2012 y enero de 2013 (Flurry, 2013).

A su vez, Flickr ha ganado terreno en el mundo de la fotografía, siendo una de las redes sociales predilectas para compartir imágenes y vídeos amateur y profesionales, como lo evidencia el crecimiento obtenido en el tercer cuarto del año 2013, con un 146 % de nuevos usuarios (Global Web Index, 2013). Esta red social, que tiene a disposición de sus usuarios un servicio de geotagging, captura los datos EXIF de la imagen o vídeo que se está subiendo al sistema y lo georreferencia en el mapa. Si los datos obtenidos no son correctos o no existen, el usuario puede cambiarlos o insertarlos a través de la plataforma y dar la ubicación exacta en la que se realizó la toma.

Ese vertiginoso crecimiento en el uso de las redes sociales y el uso de dispositivos móviles, adicionado a los esfuerzos de cada una de las compañías por georreferenciar la información generada en cada una de sus plataformas, ha propiciado un creciente interés en la ubicación de cada uno de los dispositivos, tuits y elementos de media compartidos, no sólo para tener una perspectiva espacial de los mismos, sino para obtener información útil que permita identificar nichos de usuarios para campañas de marketing, para obtener datos de vital importancia en detección y prevención de desastres, para la detección y corrección de fallas de servicio en telefonía móvil, entre otros. (Weng et al., 2011).
En este artículo se presenta un enfoque exhaustivo en la captura de información georreferenciada de las redes Twitter y Flickr, para identificar zonas de interés en el valle de Aburrá y el valle de San Nicolás, permitiendo que la información obtenida sea visible a través de herramientas como OpenStreetMaps, Google Maps y cualquier servicio de mapeo y localización.

El trabajo de Clustering se realiza en muchos aspectos de la vida cotidiana, con el fin de predecir y segmentar innumerables variables que permitan realizar desde una toma crítica de decisiones hasta una mejora sustancial entre la interacción usuario–aplicación. La utilización de esta técnica de agrupación de datos a través del análisis de sus propiedades ha permitido que se tengan importantes avances en diferentes campos de las matemáticas y las ciencias computacionales y se perciba en el mundo empresarial como una herramienta valiosa de información y obtención de ganancias.

Estudios como el realizado por Nelson & Wake (2003), que hace un breve análisis de la forma en que se adelanta la segmentación de marketing utilizando códigos postales o el de Rajagopal (2011), que utiliza técnicas de minado de datos a grandes volúmenes de información sobre clientes, muestran el gran interés que genera este tipo de clasificación y el valor real de los resultados de dichas aplicaciones.

En países como EE.UU. o Reino Unido se ha utilizado la información clínica de varios pacientes para predecir la prevalencia en ellos de algún tipo de cáncer (Cruz et al., 2006), o la utilización de diferentes biomarcadores para identificar el tipo de cáncer y aplicar la droga o tratamiento adecuado para su control y mitigación (Hijazi et al., 2012).

Por último, es posible visualizar como Kononenko (2001) realiza una buena introducción al estado del arte del machine learning en el campo de la medicina, con el objetivo de elaborar un correcto diagnóstico a los pacientes y lograr así una población más saludable y bien tratada utilizando la información que está disponible a través de diferentes fuentes.

## Metodología

*Identificación de clústeres basados en densidad y vecino cercano mejorado en las redes sociales Twitter y Flickr.*

En la minería de datos, el Density-based clustering o identificación de clústeres basados en densidad, permite la identificación de clústeres de una forma no convexa o con formas no convencionales, ya que otros algoritmos de agrupación basados en distancias, generan clústeres circulares o esféricos, lo que puede conllevar a una agrupación con altos índices de error. (Zaki et al., 2014).

En la identificación de clústeres de densidad se utiliza una estrategia que consiste en identificar regiones densas en un espacio geométrico o algebraico, separado por regiones dispersas o no densas, donde la densidad está definida como el número de elementos cercanos a otro, basados en una distancia definida previamente (Han et al., 2011). Existen diferentes algoritmos que permiten realizar dicha identificación, entre los que se destaca el algoritmo Density-based Clustering Based on Connected Regions with High Density (DBSCAN), que identifica los objetos principales que poseen vecinos densos y luego los conecta, formando una región altamente densa. Esta región identificada se conoce como clúster (Han et al., 2011). Una vez identificados los clústeres más densos, se debe pasar la información agrupada por un método de reprocesamiento que utiliza el algoritmo X-Means, que de acuerdo a sus autores se trata de una mejora al algoritmo K-Means o vecino cercano (Pelleg et al., 2000).

El algoritmo X-Means utiliza una técnica de ensamblaje, donde se generan pasos adicionales que mejoran de forma considerable los resultados obtenidos en la primera parte del proceso. Al final, estainformación generada permite a las diferentes personas encargadas del análisis de la información en una organización, presentar interpretaciones que permiten una toma de decisiones más precisa y oportuna de acuerdo a un objetivo, y en un futuro, presentar un valor agregado en cualquier cadena o espacio donde se implemente esta metodología de trabajo.

*La estructura de los tuits*

Para que sea posible identificar de forma correcta los elementos que se van a capturar y analizar, como la geolocalización de los tuits, es necesario que se conozca la estructura del objeto en cuestión. Un tuit, de acuerdo a la API de Twitter, contiene información que va desde el autor, hasta la configuración de la página del usuario distribuido en 31 módulos. La información más relevante para el análisis debe tener en cuenta el texto, la posición georreferenciada del mismo y el origen del tuit.

```
"coordinates":
{
    "coordinates":
    [
        -75.14310264,
        40.05701649
    ],
    "type":"Point"
}
```

**Figura 1.** Coordenadas en un tuit.
Fuente: Elaboración propia (2015)

```
"source":"\u003Ca href=\"http:\/\/itunes.apple.com
\/us\/app\/twitter\/id409789998?mt=12\"
rel=\"nofollow\"\u003ETwitter for Mac\u003C
\/a\u003E"
```

**Figura 2.** Origen del tuit.
Fuente: Elaboración propia (2015)

```
"text":"Tweet Button, Follow Button, and Web
Intents javascript now support SSL http:\/\/t.co
\/9fbA0oYy ^TS"
```

**Figura 3.** Texto del tuit.
Fuente: Elaboración propia (2015)

El formato de las coordenadas es el estándar GeoJSON, que indica el tipo de información y la latitud y longitud en formato decimal. El tercer bloque es el texto del tuit, que contiene los famosos 144 caracteres en formato UTF-8 y es de gran utilidad a la hora de filtrar la información para ubicar las zonas de interés sobre un tema específico.

*La estructura de la entidad de Flickr*

A diferencia de Twitter, Flickr genera entidades de tipo XML con esquemas variables que permite a un usuario obtener información diversa de una forma estructurada y comprensible. En la actualidad, la plataforma posee más de 20 puntos que posibilitan la consulta del API para extraer elementos como el número de favoritos, el número de visitas, el número de comentarios, los comentarios en sí, etc.

Para la obtención de la información de interés, primero se debe utilizar la entidad de búsqueda, que retorna el identificador de la fotografía o vídeo que contiene los datos importantes. Con el identificador de cada uno de los elementos se debe consultar si posee elementos de geolocalización. Flickr implementó un método que permite obtener estos datos de manera sencilla, entregando una segunda entidad, conocida como geo.

```xml
<photos page="2" pages="89" perpage="10" total="881">
    <photo id="2636" owner="47058503995@N01"
           secret="a123456" server="2" title="test_04"
           ispublic="1" isfriend="0" isfamily="0" />
    <photo id="2635" owner="47058503995@N01"
           secret="b123456" server="2" title="test_03"
           ispublic="0" isfriend="1" isfamily="1" />
    <photo id="2633" owner="47058503995@N01"
           secret="c123456" server="2" title="test_01"
           ispublic="1" isfriend="0" isfamily="0" />
    <photo id="2610" owner="120379497540@N01"
           secret="d123456" server="2" title="00_tall"
           ispublic="1" isfriend="0" isfamily="0" />
</photos>
```

**Figura 4.** Entidad búsqueda de Flickr.
Fuente: Elaboración propia (2015)

```
<photo id="123">
  <location latitude="-17.685895" longitude="-63.36914" accuracy="6" />
</photo>
```

**Figura 5.** Entidad geo de Flickr.
Fuente: Elaboración propia (2015)

*Almacenamiento de la información*

Una vez se conoce la estructura de la información necesaria, es importante que se genere y mantenga un repositorio donde ésta sea persistente en el tiempo. Por temas de escalabilidad (Tugores & Colet, 2013) y la divergencia en los esquemas de ambas redes sociales, se recomienda configurar la plataforma mongodb con python para garantizar su almacenamiento y posterior consulta de la misma.

El esquema de la información almacenada se divide en dos categorías:

1. Photo.
   a. Photo.geo.latitude.
   b. Photo.geo.Longitude.
   c. Photo.geo.Accuracy.
   d. Photo.name.
2. Tweet.
   a ...
   b Coordinates.
      i. Coordinates.
         1. Latitude.
         2. Longitude.
      ii. Type.
   c. Source.
   d. Text.

Así se puede garantizar la completitud de la información y es de fácil comprensión a la hora de realizar consultas en la base de datos.

**Resultados**

Existen diferentes algoritmos y técnicas conocidas como Machine Learning, que permiten entrenar a un sistema para que actúe de forma autónoma en la identificación de innumerables patrones en un volumen variable de datos (Moseley, 1988). Ésta tecnología es usada ampliamente en sectores como economía, política, criminalística, etc. Y constituye uno de los pilares fundamentales de las tecnologías emergentes como Big Data e Internet of things (Oja, 2013; Condie et al., 2013; Gershenfeld et al., 2004).

Los algoritmos de inteligencia artificial se catalogan en dos grandes grupos de acuerdo a su funcionalidad, conocidos como Inteligencia supervisada y no supervisada. El propósito del siguiente ejercicio es que se permita identificar elementos de interés sobre un grupo de datos, por lo que se utilizan técnicas de Clustering – identificación de clústeres o grupos en un conjunto de datos– en las que se incluye la identificación de clústeres basados en densidad (Kriegel et al., 2011) y en técnicas de vecino cercano mejorado, correspondientes al segundo gran grupo de esta ciencia.

Para la elaboración y comprobación del modelo, se utilizó la herramienta Weka, en su versión 3.7.1.

Estructura de la información:

| Latitud | Longitud | Texto |
|---|---|---|
| 6,2445419 | -75,6011771 | La distribución de par |
| 11,0007586 | -74,8036511 | Se vende Auteco Kim |
| 6,2445419 | -75,6011771 | FOURSQUARE: I'm at C |
| 3,4473999 | -76,5513534 | Madre e hijo se movil |
| 4,7453497 | -74,0953767 | Description: Realizad |
| 4,7439101 | -74,0967768 | Auteco Platino 125 5 S |
| 3,4473999 | -76,5513534 | El funcionario, de 30 a |
| 6,2520885 | -75,5734432 | @Dani_GallegoC san j |
| 10,9994759 | -74,804046 | Conocer personalmen |
| 10,9994634 | -74,8040359 | La planta de Auteco la |
| 10,9654431 | -74,7786864 | En AUTECO S.A.S term |
| 4,740231 | -74,0956827 | He publicado 5 fotos e |
| 4,7408823 | -74,0961583 | He publicado 13 fotos |
| 6,2520885 | -75,5734432 | La distribución de par |
| 4,7408823 | -74,0961583 | Uyyy parce ud si esta |
| 6,2520885 | -75,5734432 | Viejo busque ayuda e |
| 3,4275401 | -76,5057372 | En #JyDMotos nuestro |
| 4,7412957 | -74,0965463 | Description: Tags: Pub |
| 3,42828333 | -76,5113367 | Description: AUTECO |

**Figura 6.** Corpus normalizado de los datos.
Fuente: Elaboración propia (2015)

Al obtener información no estructurada de las redes sociales Flickr y Twitter, es necesario que se realice un proceso de normalización de la estructura a un corpus que permita al algoritmo realizar las tareas de Clustering y se pueda obtener resultados convincentes y válidos (Fraley & Raftery. 2002). Para ello se sugiere utilizar los valores de longitud y latitud, y como referencia un tercer campo que permite identificar el tuit o el nombre de la fotografía una vez se hayan aplicado las agrupaciones correspondientes.

La longitud y la latitud a procesar deben estar en formato decimal –representación en base 10-, permitiendo realizar operaciones matemáticas sobre los valores que son útiles en otras técnicas, como regresión numérica y logística.

*Entrenamiento del algoritmo*

En la creación y utilización de modelos de Machine learning es necesario realizar un proceso de entrenamiento al algoritmo autónomo, con el fin de identificar patrones en los datos que permiten obtener parámetros, los cuales se utilizan para predecir el estado final de los conjuntos de datos que se proporcionan a futuro.

En este ejercicio se realizó una recolección de 20.495 tuits y 16.958 fotografías únicas, correspondientes al área metropolitana del valle de Aburrá y al valle de San Nicolás, en el departamento de Antioquia, Colombia. Los términos usados en la búsqueda de la información fueron Medellín, Fiesta y 4sq.com, permitiendo que se pueda realizar la identificación de las 10 zonas de interés para posibles turistas que deseen visitar la capital antioqueña y los cascos urbanos aledaños al área metropolitana.

El algoritmo X-Means, también conocido como algoritmo de vecino cercano mejorado, identifica de manera dinámica el número de clústeres y sus respectivos centroides, indicándoles parámetros mínimos y máximos de agrupaciones deseadas. Un centroide es definido como "un punto de un conjunto Є, que tiende a tener la propiedad de ser el centro de una cobertura esférica con el mínimo radio posible" (MacQueen, 1967). Es decir, representa el elemento más central de un conjunto de datos Є, donde la distancia a cualquier punto A que pertenece a Є, es equivalente a la media de todas las posibles distancias dentro del mismo.

A continuación, se deben generar 10 clústeres – mínimo 10, máximo 10- en la configuración del algoritmo X-Means, con el fin de obtener 10 puntos centroides calculados y poderlos graficar en el mapa. El resultado es el siguiente:

```
Cluster centers            : 10 centers
Cluster 0
        6.152541316281556 -75.35414110567953
Cluster 1
        7.538536774017089 -74.96247213102559
Cluster 2
        6.241243759319632 -75.57945209898037
Cluster 3
        7.930696282857142 -76.6294891260317
Cluster 4
        6.054656377323939 -75.47818354355634
Cluster 5
        6.3988872879655725 -75.486975038795001
Cluster 6
        6.189913774778162 -75.58028185518423
Cluster 7
        6.169138582367698 -75.62828644540573
Cluster 8
        6.068931923758866 -75.6605151760283
Cluster 9
        6.3144177775225145 -75.57233013551647
```

**Figura 7.** Centroides del grupo de clústeres.
Fuente: Elaboración propia (2015)

Como es visible en la Figura 7, el algoritmo de clústeres basados en densidad realizó un cómputo de cada uno de los 10 grupos identificados por el algoritmo X-Means y con cada uno de sus centroides calculados, de acuerdo a la densidad de cada agrupación. La información muestra el número de centros obtenidos (10 centers), el número del clúster en forma 0-Index y su respectivo centroide con forma Latitud, Longitud.

Si se realiza un gráfico de estos puntos en un mapa, se obtiene lo siguiente:

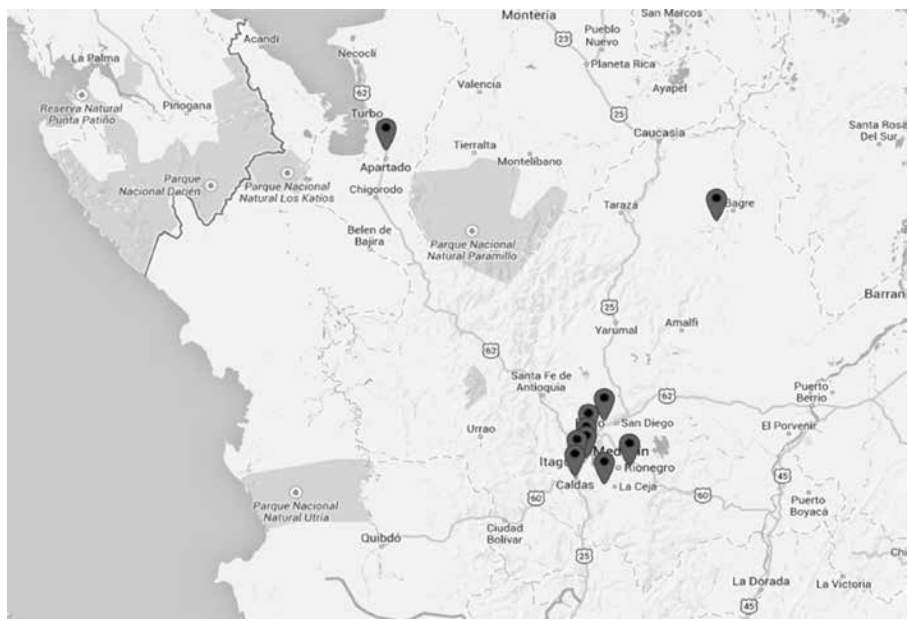

**Figura 8.** Visión de los 10 clústeres.
Fuente: Elaboración propia (2015)

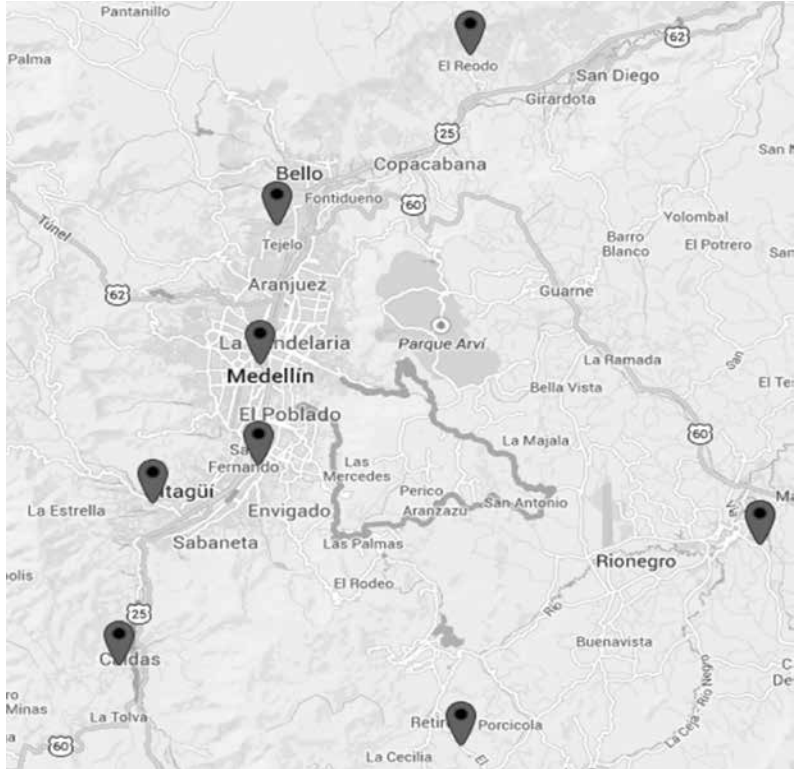

**Figura 9.** Clústeres que cubren el valle de Aburrá y el valle de San Nicolás.
Fuente: Elaboración propia (2015)

Gracias al proceso de Clustering, es posible identificar elementos que no pertenecen a la zona de interés para descartarlos. Al parecer, es un bug que presenta la API de Twitter con el Streaming API.

En la Figura 9 se pueden observar los clústeres restantes ubicados en la zona deseada: el valle de Aburrá y el valle de San Nicolás. Los centroides, al ser calculados por el método utilizado, representan el conjunto de las palabras clave y no una de las palabras seleccionadas.

*Identificación del radio de cobertura del clúster*

Gracias al formato decimal de los elementos latitud y longitud, es posible conocer el radio de cobertura que posee un clúster en la información, donde el radio de cobertura se define como la distancia existente entre el punto formado por las medias de cada una de las latitudes y longitudes del clúster –que se denominará punto de medias-, y el punto con mayor latitud y longitud del mismo conjunto. La circunferencia resultante, llamada circunferencia de cobertura, tendrá como centro el centroide del clúster deseado.

Para realizar el cálculo del radio de cobertura se utiliza el método de la Fórmula Haversine (Sinnott, 1984), planteado en 1984 por un integrante de la NASA y expresada en la fórmula (1):

$$sea\ haversin(\theta) = \sin^2\frac{\theta}{2} = \frac{1-\cos\theta}{2} \quad (1)$$

De lo anterior es posible definir para cualquier pareja de duplas (latitud, longitud), expresados en radianes, lo siguiente:

$$\text{harvesin}\left(\frac{d}{R}\right) = \text{harvesin}(\mathcal{L}_1 - \mathcal{L}_2) + \left(\cos\mathcal{L}_1 * \cos\mathcal{L}_2 * \text{harvesin}(\Delta x)\right) \quad (2)$$

Donde:
d: distancia entre puntos en kms
L_1: latitud del punto de medias
L_2: latitud del punto más lejano
R: Radio de la tierra
Δx: Diferencia entre longitudes de los puntos

De la fórmula (2) se puede concluir, utilizando matemática elemental y asumiendo que la función haversin(x) es biyectiva, que:

$$d = R\,\text{haversin}^{-1}(h) = 2R \arcsin\left(\sqrt{h}\right) \quad (3)$$

$$\text{donde } h = \text{haversin}\left(\frac{d}{R}\right) \quad (3.1)$$

A continuación, se debe aplicar el método propuesto en la fórmula (3). Para el cálculo de distancias en dos clústeres (clústeres dos y seis), al operar con cada uno de los grupos para la identificación del radio de cobertura, se obtiene la siguiente información:

**Tabla 1.** Información sobre el clúster número seis.

| Elementos | Valores |
|---|---|
| Clúster | Clúster número seis |
| Punto distante | $(-75.49676, 6.354782)$ |
| Punto de medias | $(-75.58002, 6.18991)$ |
| Centroide | $(-75.58028, 6.189913)$ |
| Radio de cobertura | $2.855\ k\ m\ s$ |

Fuente: Elaboración propia (2015).

**Tabla 2.** Información sobre el clúster número dos.

| Elementos | Valores |
|---|---|
| Clúster | Clúster número dos |
| Punto distante | $(-75.57941, 6,273949)$ |
| Punto de medias | $(-75.5795, 6.2412)$ |
| Centroide | $(-75.5794, 6.2412437)$ |
| Radio de cobertura | $3.622\ k\ m\ s$ |

Fuente: Elaboración propia (2015).

Si se utiliza la información especificada en las Tabla 1 y Tabla 2, es posible realizar una visualización como las que se muestran en las figuras 10 y 11, dibujando los círculos de cobertura, definidos por el radio de cobertura calculado y el centroide del clúster correspondiente.

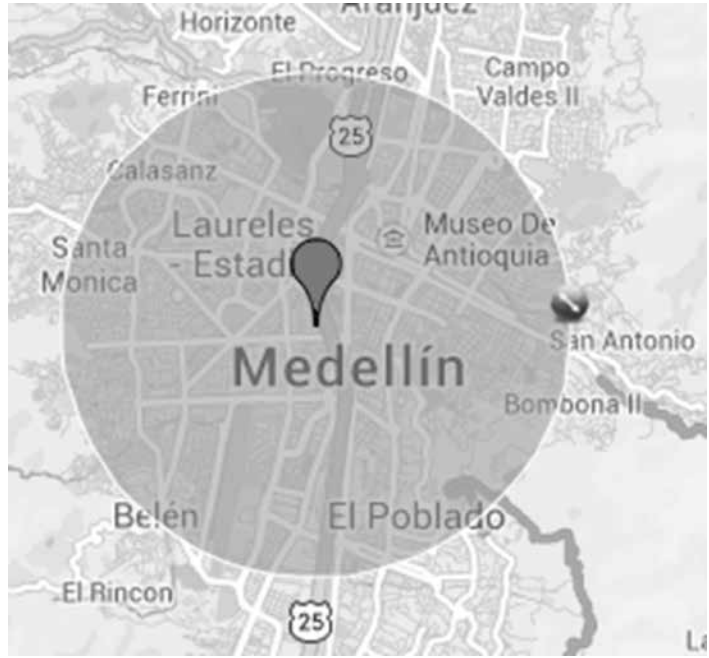

**Figura 10.** Circunferencia de cobertura del clúster dos, con radio de 3.622 kms
Fuente: Elaboración propia (2015)

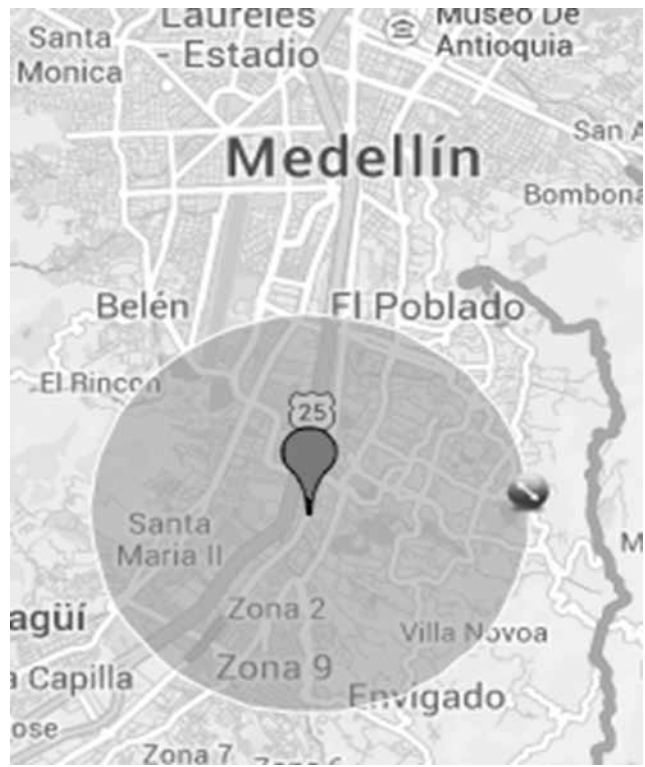

**Figura 11.** Circunferencia de cobertura del clúster seis, con radio 2.855 kms
Fuente: Elaboración propia (2015)

*Análisis de la información obtenida por clústeres*

Al revisar, la información correspondiente a cada uno de los clústeres analizados, es posible evidenciar un agrupamiento de la palabra clave 4sq.com en el clúster seis, que posee una circunferencia de influencia ubicado en la zona rosa de la ciudad de Medellín y Envigado. La palabra Medellín está ubicada de forma mayoritaria en el clúster dos y la palabra Fiesta, está combinada en ambos clústeres. Vale afirmar que aunque la agrupación de la información se realiza utilizando los elementos de latitud, longitud y texto de cada uno de los datos, uno o varios puntos pueden encontrarse ubicados dentro de la circunferencia de influencia de cada uno de los clústeres, como se demuestra en los gráficos de dispersión (figuras 12 y 13); el clúster dos es representado en color negro y el clúster seis en color amarillo.

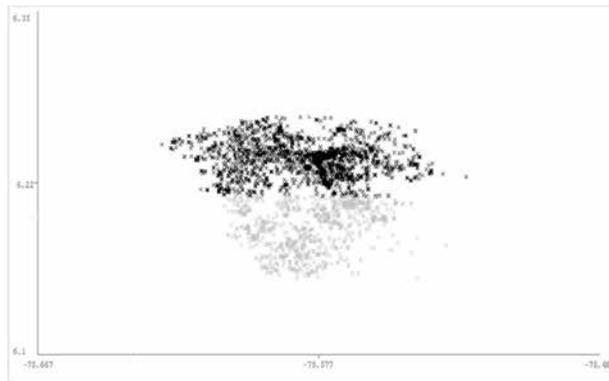

**Figura 12.** Gráfico de dispersión clústeres dos y seis
Fuente: Elaboración propia (2015)

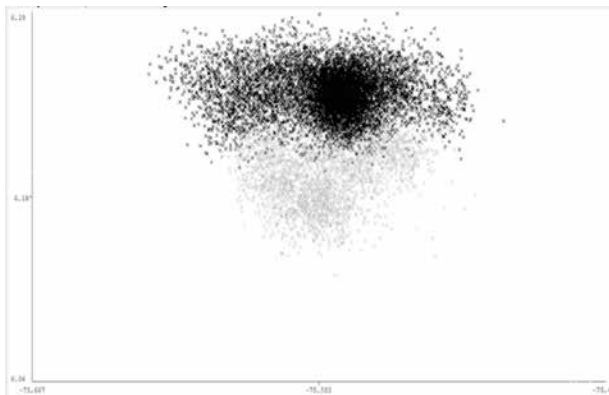

**Figura 13.** Intersección entre el clúster dos y seis
Fuente: Elaboración propia (2015)

**Conclusiones**

Sin duda alguna, las técnicas de Machine Learning comprenden una serie de soluciones óptimas a diferentes problemas y necesidades, producidos por el surgimiento de las redes sociales virtuales y el boom de Internet. La segmentación, en éste y en otro sinfín de casos, es crucial para obtener unidades de valor fundamentales para el negocio o la idea que se desea analizar y, sin duda alguna, el método expuesto en este artículo responde a esos interrogantes de una manera relativamente ágil y sencilla.

En la actualidad se está trabajando en un sistema inteligente que actúe en tiempo real, realizando tareas de entrenamiento y evaluación de manera automática y transparente para los usuarios finales, y la construcción de informes y visualizaciones con unos pocos clics a partir de información de diversas fuentes como Twitter, Flickr, Instagram y otros.

El producto final podrá ser muy útil para casas de periodismo, entidades gubernamentales y entidades de atención al desastre para poder entregar un servicio más rápido y eficiente.

**Agradecimientos**

Es de vital importancia agradecer al profesor Ing. Jorge M. Gaviria H. por su dedicación, enseñanza y apoyo en este proyecto.

**Referencias**

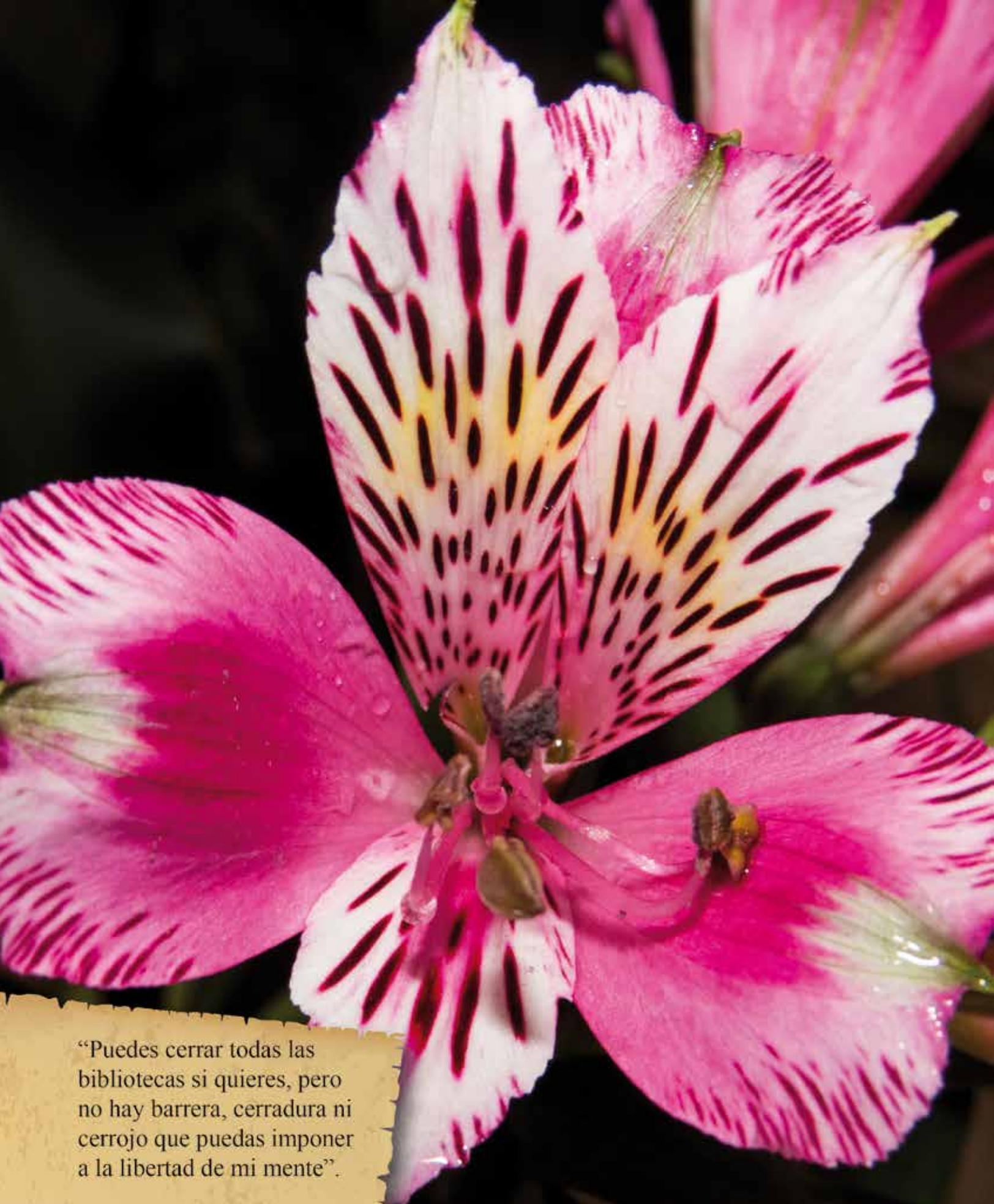

"Puedes cerrar todas las bibliotecas si quieres, pero no hay barrera, cerradura ni cerrojo que puedas imponer a la libertad de mi mente".

Virginia Woolf.

Alstroemeria /Autor: Diego Alonso Rivera Vergara